\documentclass{iaus}
\usepackage{graphicx}

\def\mso{M_\odot}

\def\simgr{\,\hbox{\hbox{$ > $}\kern -0.8em \lower 1.0ex\hbox{$\sim$}}\,}
\def\simle{\,\hbox{\hbox{$ < $}\kern -0.8em \lower 1.0ex\hbox{$\sim$}}\,}

\title[Boron in massive stars] 
{Light elements in massive single and binary stars }

\author[N. Langer et al.]   
{N. Langer$^{1,2}$, I. Brott$^2$, M. Cantiello$^2$, S.E. de Mink$^2$, R.G. Izzard$^3$, 
 \and S.-C. Yoon$^1$}

\affiliation{
$^1$Argelander-Institut f\"ur Astronomie, Universit\"at Bonn, \\
Auf dem H\"ugel 71, Germany  \\[\affilskip]
$^2$Sterrenkundig Instituut, University of Utrecht, \\ Postbus 80000,
NL-3508TA, Utrecht, the Netherlands  \\[\affilskip]
$^3$Institut d'Astronomie et d'Astrophysique, Universite Libre de Bruxelles, \\
Boulevard du Triomphe, 1050 Brussels
}

\pubyear{2010}
\volume{268}  
\jname{Light elements in the Universe}
\editors{C. Charbonnel, M. Tosi, F. Primas \& C. Chiappini, eds.}
\begin{document}

\maketitle

\begin{abstract}
We highlight the role of the light elements (Li, Be, B) in the evolution of massive single and
binary stars, which is largely restricted to a diagnostic value, and foremost so for the element
boron. However, we show that the boron surface abundance in massive early type stars
contains key information about their foregoing evolution which is not obtainable otherwise.
In particular, it allows to constrain internal mixing processes and potential previous
mass transfer event for binary stars (even if the companion has disappeared).
It may also help solving the mystery of the slowly rotating nitrogen-rich massive main
sequence stars. 
\end{abstract}

\firstsection 
\section{Introduction}

A large effort has been undertaken in the last decades to measure and understand the
surface chemical composition of massive main sequence stars. In particular, the detection
of nitrogen enhancements in quite a number of such stars (e.g., Gies \&
Lambert 1992) has triggered the idea that internal mixing processes can bring
material from the stellar core to the surface in rapid rotators
(Meynet \& Maeder 2000, Heger and Langer 2000). 

The picture has become more complicated by the recent analysis of a large sample
of early B~type  main sequence stars of Hunter et al. (2008), who showed that
the nitrogen-rich stars found by Gies \& Lambert (1992), who restricted their
analysis to objects with low projected rotational velocities, 
are likely part of a population of {\em intrinsically} slowly rotation main sequence
stars. This view is supported by the work of Morel et al. (2006, 2008), who indeed
identifies such a population in our Galaxy (see also Morel, 2009). 
The origin of the nitrogen enrichment in these stars is not understood,
but as they are slow rotators it appears difficult to reconcile them with
the idea of rotational mixing.

On the other hand, Hunter et al. (2008) also identified a nitrogen-rich population
of rapidly rotating early B stars, which appears to be well in line with the predictions
of theoretical models including rotational mixing (cf., Maeder et al. 2008).
The caveat here is that evolutionary models of massive close binaries ---
whether or not they include rotationally induced chemical mixing (Langer et al. 1998) ---
appear to predict essentially the same trend of nitrogen enrichment
with rotational velocity as the single star models (Meynet \& Maeder 2000, Heger and Langer 2000).

The light elements lithium, beryllium and boron may play a key role
to resolve this issue. They are so rare in the interstellar
medium that they can not influence the course of stellar evolution,
and thus are often neglected in massive star models. 
Furthermore, they are fragile nuclei which are generally not synthesised in stars,
but rather destroyed, at least certainly on the main sequence.
In the cool low mass stars, lithium is most interesting, as it can be observed rather
easily, and it is indeed used extensively to constrain internal mixing processes
as can be seen in many contributions to this book.
In massive stars this role can be played by boron as will be outlined below.

\section{Single stars}

\begin{figure}[b]
\begin{center}
 \includegraphics[width=4.8in]{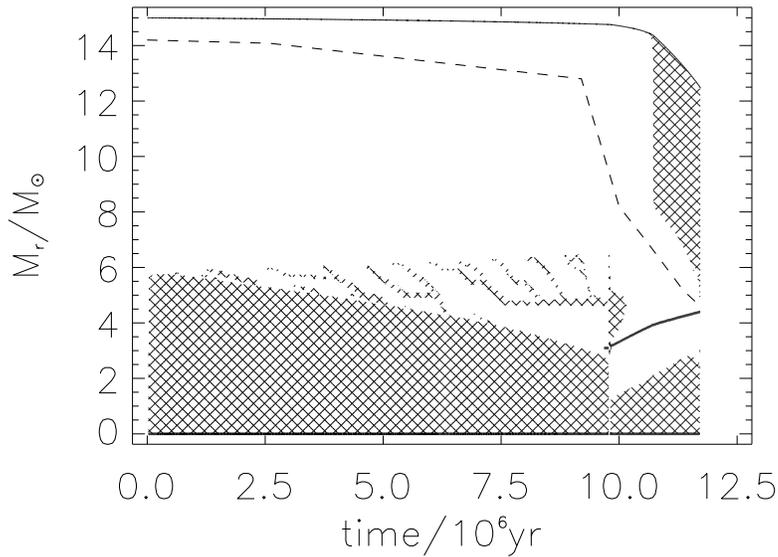} 
 \caption{Internal structure of a 15$\mso$ star during core hydrogen
and helium burning. The solid line on top indicates the total mass
of the star as function of time. Hatched areas designate convectively
unstable mass zones in the star. The full drawn line at $M_r\simeq 4\mso$
and $t \simgr 10^7\,$yr designates the location of the H-burning shell
during core helium burning. The dashed line indicates the threshold
temperature for boron destruction (cf. Fliegner et al. 1996).}
\end{center}
\end{figure}

\begin{figure}[t]
\begin{center}
 \includegraphics[width=3.8in, angle=270]{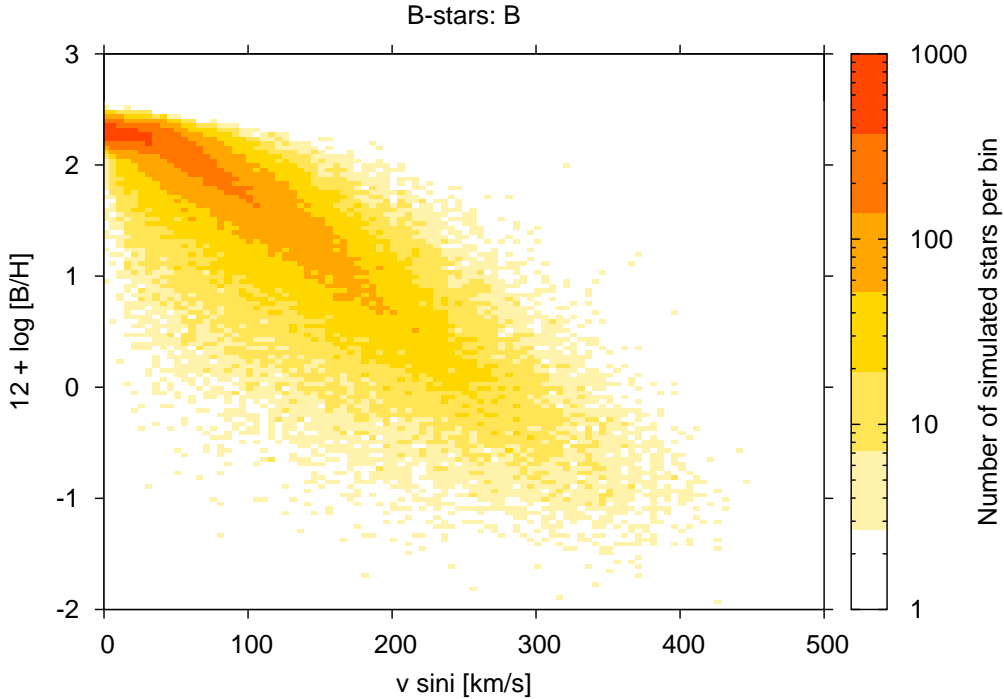}
 \caption{Result of a population synthesis calculation for massive main sequence
 stars (Brott et al., in prep.), employing a Salpeter initial mass function,
 and a distribution of initial stellar rotation rates as derived by Hunter et al. (2008a),
 and a constant star formation rate,
 based on single star evolution models which include rotational mixing.
 A random Gaussian error of 0.2 dex was added to the predicted boron abundances.
 }
\end{center}
\end{figure}

\begin{figure}[t]
\begin{center}
 \includegraphics[width=3.9in]{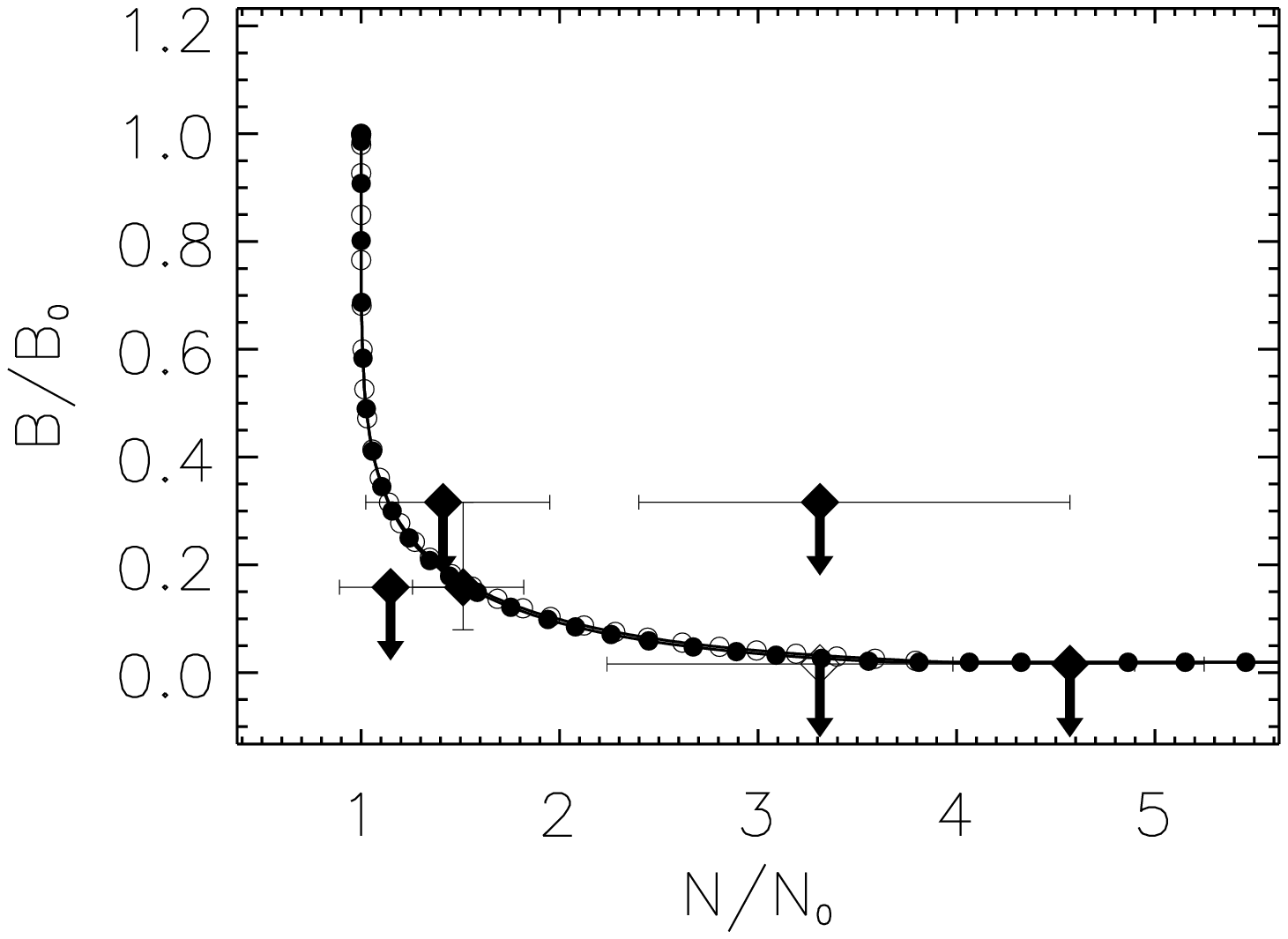}
 \caption{Boron depletion versus nitrogen enrichment for
rotating stars of $10\mso$
during the main sequence evolution (Fliegner et al. 1996). Open circles correspond to a
moderately, filled circles to a fast rotating model. Two
neighbouring symbols on each line indicate a constant
distance in time of $10^6$~yr.
Diamonds indicate the location of the five B main sequence stars
(filled symbols) and one B~giant (open symbol)
of Venn et al.'s (1996) sample (see also Venn et al. 2002). Arrows designate upper limits.
}
\end{center}
\end{figure}

Other than helium and the major CNO nuclei,
the light elements Li, Be, and B are destroyed by proton capture 
relatively close to the stellar surface.
For both stable boron isotopes, $^{10}$B and $^{11}$B,
the life time against proton capture is equal to the main sequence life time of a 10$\mso$ star
(10$^7 \,$yr) at a temperature of roughly
$7\, 10^6\,$K.    
Fliegner et al. (1996) have computed the evolution of stars of 15$\mso$, and found
this temperature to occur sufficiently
deep inside the stellar envelope (i.e. roughly 1$\mso$ below the
surface; cf. Fig.~1) that its surface abundance can not be altered due
to mass loss alone on the main sequence in the B~star regime. Thus, the
boron abundance in B~stars is a critical test of mixing processes
in the upper stellar envelope, while CNO and helium abundances additionally
trace the mixing in deeper layers.

Models which include rotational mixing show that boron depletion at the stellar
surface is predicted for initial rotational velocities above 50~km/s. 
However, while in low mass stars, the rotational velocity is a strong function
of age, and so is the lithium abundance, a clear correlation of boron with
stellar age is not expected in a population which contains initially fast and
slow rotators (cf. Gies \& Lambert 1992).  
However, a population synthesis simulation by Ines Brott shows (Fig. 2), that
a clear correlation of the surface boron depletion with the stellar rotation
rate can be expected in early B~type single stars. 

Figure~3 shows the behaviour of the surface abundances of a rotating 15$\mso$
in a plot of boron depletion versus nitrogen enhancement.
Interestingly, it predicts that boron depletion happens essentially
{\em before} nitrogen enrichment occurs. Thus, stars which are already depleted
in boron, but which are still nitrogen normal are expected. Indeed, in the sample
of Venn et al. (1996), several such stars seem to exist. The importance of this
finding becomes more clear in the next section.

\section{Binary stars}

\begin{figure}[t]
\begin{center}
 \includegraphics[width=\textwidth]{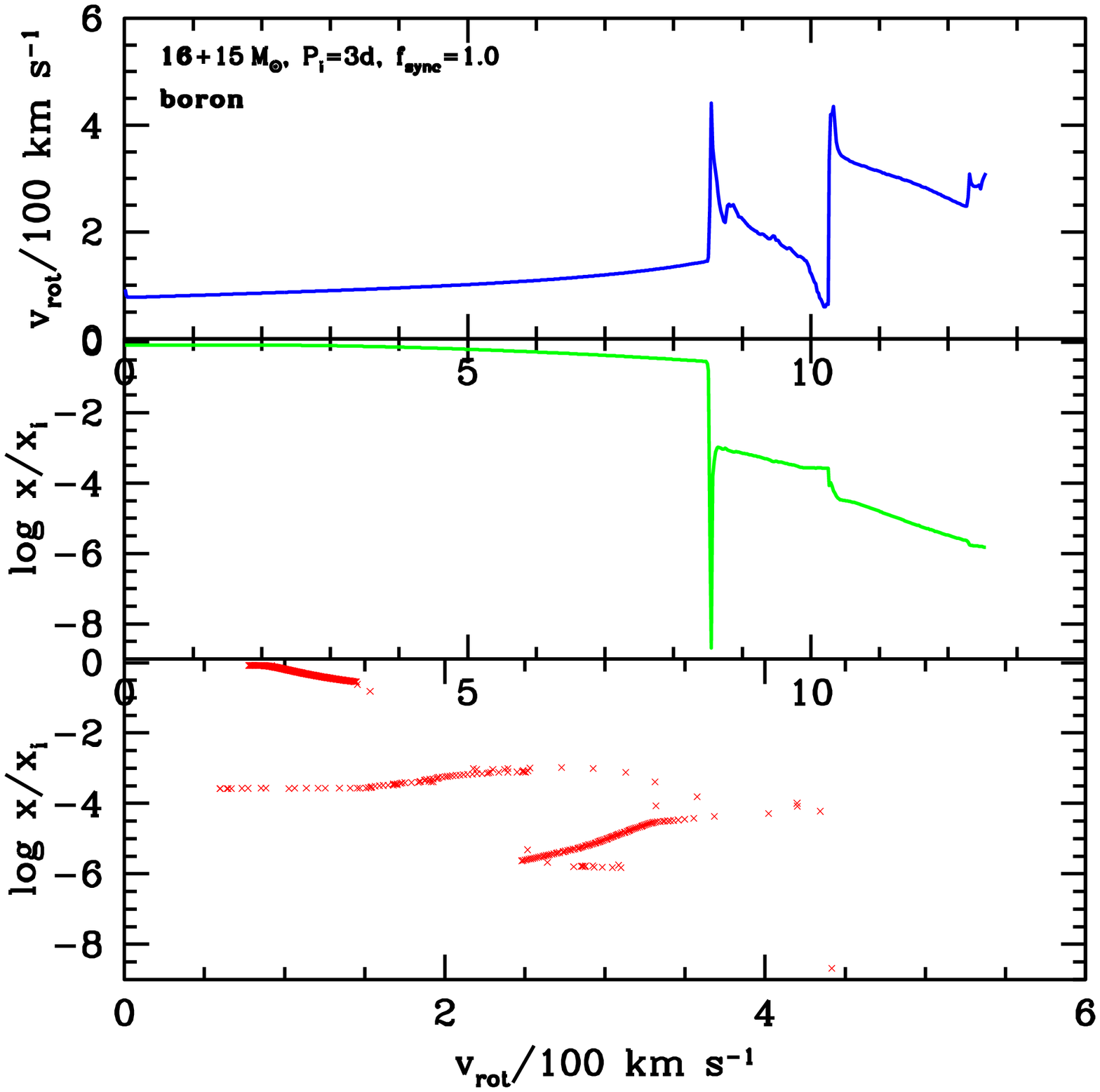}
 \caption{Equatorial rotational velocity (upper panel) and
surface boron mass fraction relative to the initial value
(middle panel) as function of time (in Myr), for the mass gainer in a
solar metallicity
$16\mso + 15\mso$ binary with an initial orbital period of 3~days.
The computations include the physics of rotation for both
components as in Heger et al. (2000), and Spin-Orbit coupling
as in Detmers et al. (2008) with the nominal coupling parameter
$f_{\rm sync}=1$, and rotationally enhanced stellar
wind mass loss (Langer 1998). Internal magnetic fields are not
included. The bottom panel shows the evolution of the mass gainer
in the boron depletion versus rotational velocity diagram,
where each data point represents a duration of 20\,000\,yr.
The spin-down of the star after the first accretion event
($t=8.5...10\,$Myr) is mostly due to tidal effects.
}
\end{center}
\end{figure}

\begin{figure}[t]
\begin{center}
 \includegraphics[width=\textwidth]{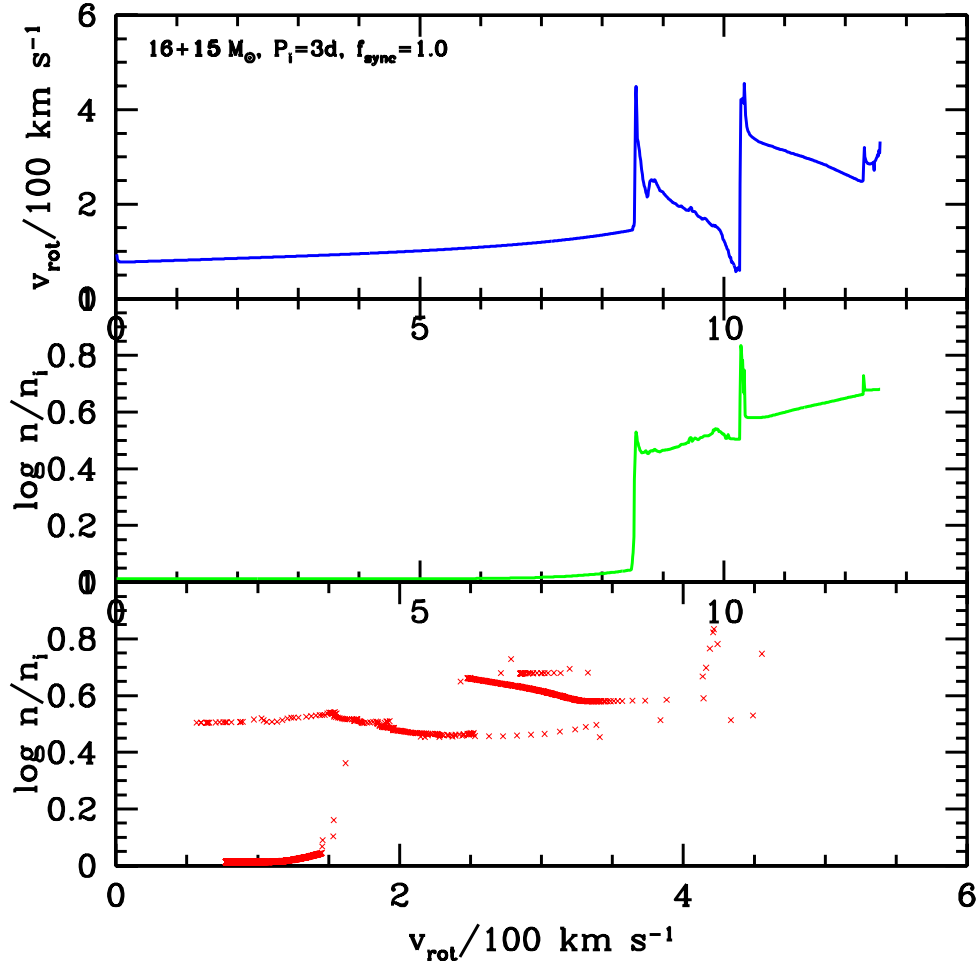}
 \caption{As Fig.~4, but showing the surface nitrogen abundance.}
\end{center}
\end{figure}

\begin{figure}[t]
\begin{center}
 \includegraphics[width=0.48\textwidth]{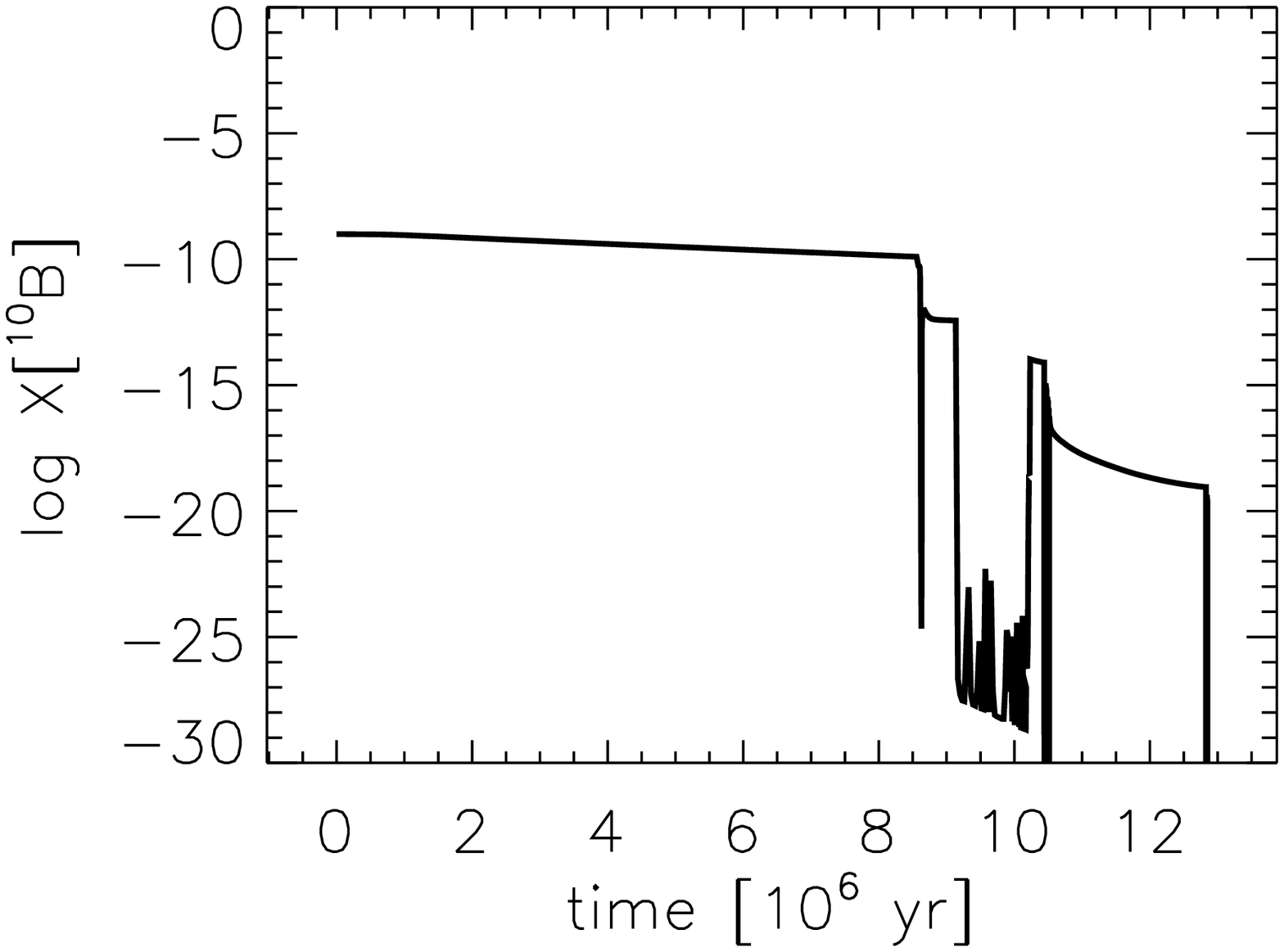}
 \includegraphics[width=0.48\textwidth]{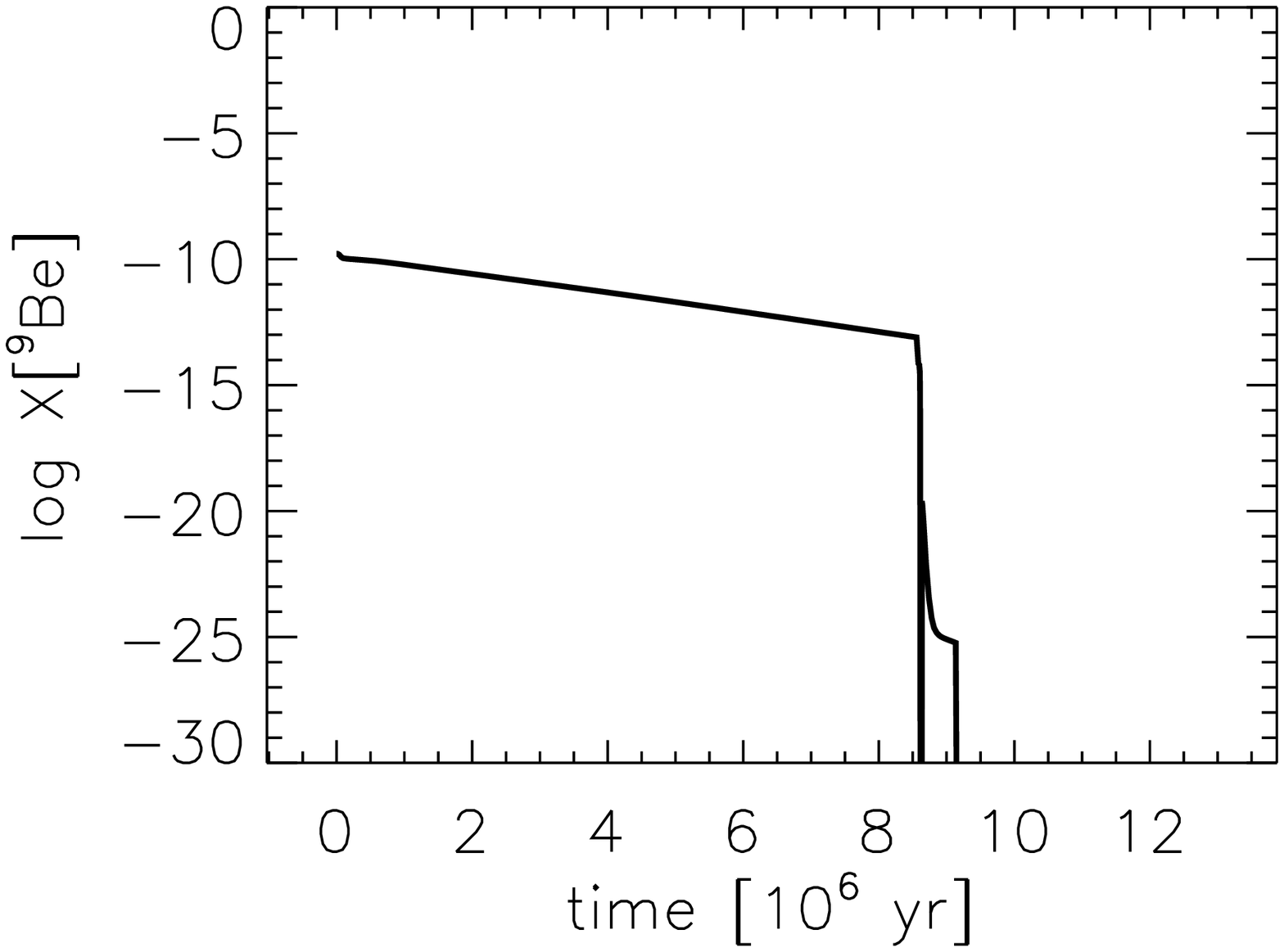}
 \caption{Surface abundance of boron~10 (left) and of beryllium (right) as
function of time during cure hydrogen burning, for the mass gainer
in a solar metallicity $16\mso + 15\mso$ binary with an initial period of 3~days.
In contrast to the models shown in Figs.~4 and~5, the stellar models shown here have been
computed including angular momentum transport by internal magnetic fields
(Yoon et al., in prep.). }
\end{center}
\end{figure}

The binary fraction of massive stars is very high, and it may be futile to
try to understand their surface abundances without considering effects of binarity
(Langer et al. 2008). The issue of binarity is not easily resolved, because
of two reasons. 

Firstly, after a strong binary interaction, the object may not
appear to be a binary any more. This is so since many mass transfer systems
will produce a rejuvenated main sequence star which dominates the light of
the system, with a faint helium star in a wide orbit. In fact, many such objects
ought to exist, as we see many of their descendants, the Be-X-ray binaries.
However, for the stage where the main sequence star has a helium star companion,
we practically do not know any counterpart. Also, there will be many cases where
the post-interaction system is in fact a single star. E.g., in many 
systems consisting of a
main sequence star and a helium star companion, the supernova explosion of the
helium star will disrupt the binary (instead of leading to a Be-X-ray binary stage).
And furthermore, as many as 10\% of all massive stars may actually merge with their
close companion during the main sequence evolution. 
Therefore, most of the binaries which we detect in massive main sequence star
populations may have in fact not yet interacted, while on the other hand,
many apparent single stars may be the result of a strong binary interaction.

Secondly, binary interaction has many complex branches, and it is a big theoretical
enterprise to even fully consider the most important ones in a population study.
So far, no such study exists which takes the physics of rotation fully into
account --- which is what appears to be needed in order to demonstrate that rotational
mixing works in Nature.

However, there are some binary evolution models available which include all the required
physics, and while they will not allow to obtain a full view of the picture, they may give
indications in one or the other direction. To that purpose, Fig.~4 shows the time evolution
of the surface rotational velocity and of the surface boron abundance of the mass gainer
(i.e. the star which will be visible after the mass transfer) of a close massive binary.
The lower most panel also shows the boron depletion factor as function of the rotational
velocity in this star.

The model shown in Fig.~4 does include rotational mixing, which produces the mild
boron depletion before the first mass transfer event (at about $t=8.5$ Myr).
The mass transfer event itself, which occurs on a time scale of some 10$^4$\,yr,
puts boron depleted layers on the surface of the mass gainer, and subsequent
thermohaline mixing brings the surface boron abundance back to a level of one per mille
of the initial boron abundance. 

Consecutively, the mass gainer is spun down by tidal interaction, while rotational mixing
reduces the surface boron abundance slightly more. As a result of this phase,
one obvious result from Fig.~4 is that in mass gainers of close binaries, 
one can {\em not} generally expect a correlation of the boron depletion factor
with the surface rotation rate. Here, a slowly rotating main sequence star is produced
which shows a surface boron depletion by 3 to 4 orders of magnitude. Fig.~2 shows
that such a strong depletion is only expected in single stars with rotation rates
above 300 km/s. 

The model in Fig.~4 suffers from a second mass transfer at $t\simeq 10.2\,$Myr, which
leads to a second spin-up of the mass gainer and a further strong reduction of the
surface boron abundance. As the orbit widens strongly during the second mass transfer
phase, no tidal spin-down occurs thereafter. The mass gainer is now a rapidly rotating
main sequence star which is strongly boron depleted. 

While rotational mixing is included in this model, we want to point out that the effects of
rotational mixing and of mass transfer (and thermohaline mixing) are well separable in Fig.~4.
The steps in the time evolution of the boron abundance are produced by the mass transfer,
and the long time scale changes are due to rotational mixing. We conclude that the main
features of the boron evolution of this model would not change if rotational mixing were
switched off. From this consideration, we can argue that if rotational mixing would not
operate in Nature, then perhaps binary models would not predict main sequence stars
with only a mild boron depletion (unless one considered masses much higher than 15$\mso$,
where mass loss can gradually uncover boron-depleted layers).

The situation becomes more clear when nitrogen is considered at the same time. Fig.~5
shows the nitrogen surface abundances of the same binary model described above. 
The nitrogen surface abundance increases abruptly due to the mass transfer. Thus, from this
sequence, one would not expect to observe stars which are boron depleted but not
nitrogen enriched unless rotational mixing operates. The corresponding stars
in Fig.~2 thus indicate that rotational mixing is perhaps operating as we expect.

However, this is not yet a definite conclusion. The example binary displayed in Figs.~4
and~5 evolves rather conservatively, i.e. about 70\% of the transferred matter 
is actually accepted by the mass gainer (while the rest is ejected due to its excess
angular momentum). We know from observations and predict
theoretically that many mass transfer systems evolve rather non-conservatively
(Petrovic et al. 2005ab). We can currently not exclude that some of these systems
produce boron depletion with only mild or no nitrogen enhancement.

It is also interesting to consider boron as a test of rotational mixing in
very close and very massive pre-mass transfer binaries, as explored by de Mink et al. (2009).
In such very tight binaries, tidal synchronisation can enforce very rapid rotation
of both stars. While de Mink et al. point out that generally the nitrogen surface abundance
is the prime observable for such test, their results on boron depletion  
appear particularly interesting for Galactic binaries, since in those 
boron has a larger predicted relative change then nitrogen.

Finally, in Fig.~6, we show the surface abundances of boron and of beryllium in a
similar binary as the one discussed above (even though some physics assumptions were
different, which is not essential for our discussion here).
Fig.~6 indicates that also beryllium is very interesting from the theoretical
point of view. However, beryllium abundance determinations in hot main sequence
stars appear to be difficult.

\section{Unknown mixing processes}

\begin{figure}[t]
\begin{center}
 \includegraphics[width=4.4in]{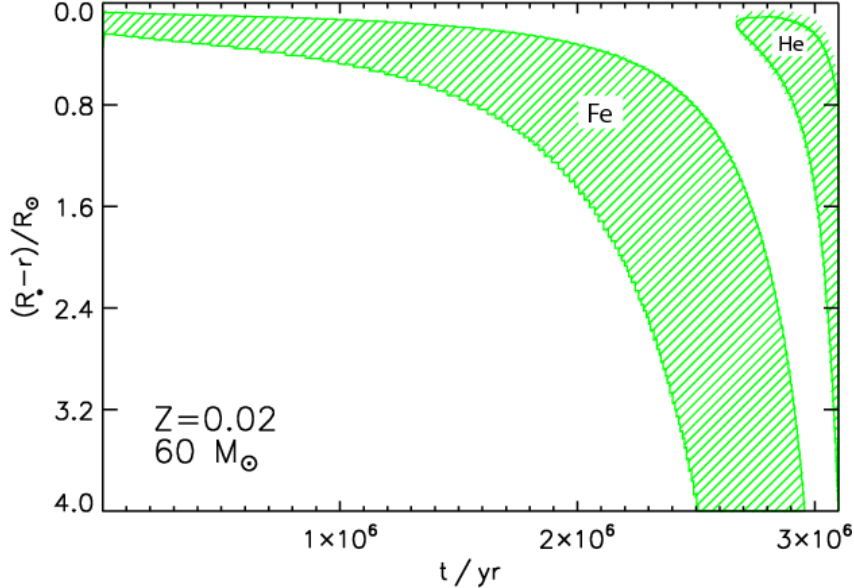}
 \caption{
Evolution of the radial extent of the subsurface helium and iron convective regions (hatched) 
as function of time, from the zero age main sequence to roughly the end of core hydrogen burning, 
for a 60  $~{M}_\odot$ star (Cantiello et al. 2009). The top of the plot represents the stellar surface. Only the upper 
4$~{R}_\odot$ of the star are shown in the plot, while the stellar radius itself increases 
during the evolution. The star has a metallicity of Z=0.02, and its effective temperature decreases 
from 48 000 K to 18 000 K during the main sequence phase.}
\end{center}
\end{figure}

\begin{figure}[t]
\begin{center}
 \includegraphics[width=\textwidth]{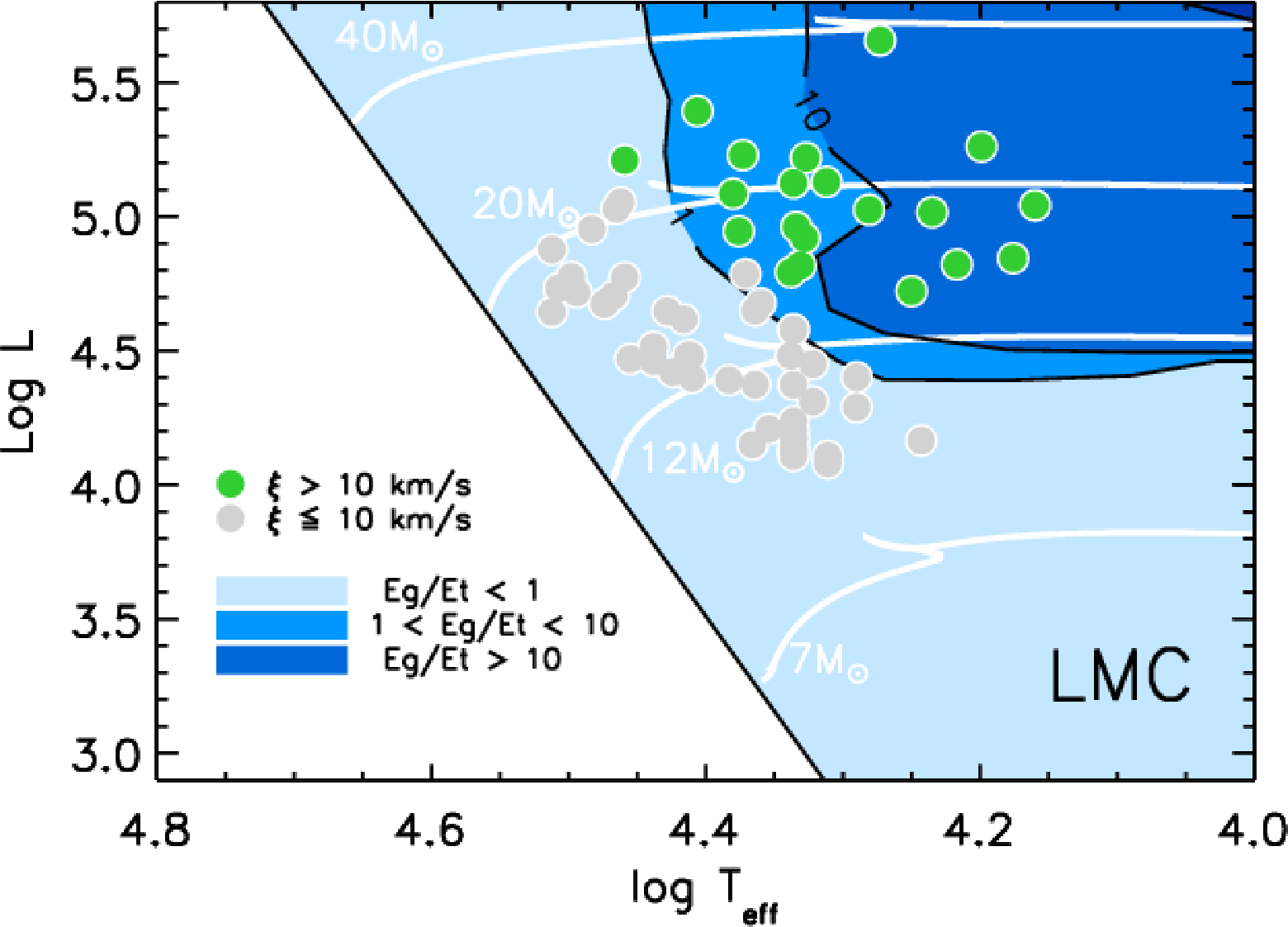}
 \caption{Values of the ratio $E_{{\rm g}}/E_{\rm s}$ of the kinetic energy in the form of 
gravity waves above the iron convection zone, to the kinetic energy of the surface velocity field, 
as a function of the location in the HR diagram (Cantiello et al. 2009). This plot is based on evolutionary 
models between 5 $~{M}_\odot$ and 100 $~{M}_\odot$ for LMC metallicity. The ratio 
$E_{{\rm g}}/E_{\rm s}$ (see Eq. (9) of Cantiello et al. 2009) is estimated using a value $v_{\rm s} = 10~ {\rm km}~ {\rm s}^{-1}$ 
for the surface velocity amplitude. Over-plotted as filled circles are stars which have photospheric 
microturbulent velocities $\xi $ derived in a consistent way by Hunter et al. (2008a). 
Only data for stars with an apparent rotational velocity of $v\sin i< 80~ {\rm km}~ {\rm s}^{-1}$ is plotted. 
Solid white lines are reference evolutionary tracks, and the full drawn black line corresponds to the zero age main sequence. }
\end{center}
\end{figure}

We have seen in the previous sections, that the surface abundances of massive main sequence stars
are not yet fully understood, and that unambiguous evidence for the existence of rotationally induced
mixing in massive stars is still lacking. Of particular worry is the solid evidence for a
population of nitrogen-rich slowly rotating massive main sequence stars (Hunter et al. 2008, 2009;
Morel et al. 2006, 2008; Morel 2009). While the binary models discussed above do show a way
to produce such stars (Langer et al., 2008; cf. Fig.~5), the Galactic fraction of this
population contains well investigated $\beta$\,Cephei pulsators (Morel et al. 2006, 2008)
none of which seems to show any indication of binarity. Despite the warning above that
binarity might be difficult to detect in post-mass transfer systems, the lack of any indication
of a companion in well studied nearby stars could imply that binarity is not the (only)
answer to this question. It also remains to be seen whether binary evolution could produce
enough of these objects, which may have a frequency of about 15\% of all main sequence stars
(Hunter et al. 2008). 

So we may face the situation that so far completely unaccounted mixing processes may operate
in stars (see also Brott et al., in preparation). 
Perhaps, they could be related to magnetic fields in the interior of massive stars.
Also gravity waves could be excited in massive stars, as Talon \& Charbonnel (2008) employ
them for angular momentum transport in intermediate mass stars. However, we want to end with 
a related possibility, for which recent observational evidence has accumulated.

Cantiello et al. (2009) investigated the subsurface convection zones which occur in the envelopes
of hot massive stars due to the iron opacity peak (cf. Fig.~7).
They found that the occurrence of strong subsurface convective motion predicted
by the models correlates with observed large microturbulent velocities deduced from
stellar spectroscopy. While this may not be sufficient to conclude that 
subsurface convection causes observable motion at the stellar surface, it appears 
to be a possibility. This could mean that the subsurface layers of massive stars,
independent of their rotation, could be in motion, and perhaps lead to some mixing
near the surface; this might produce a surface boron depletion without changing
nitrogen. 

\section{Conclusions}

It remains a major challenge to the theory of massive star evolution to explain
the observed surface abundances of massive main sequence stars.
While until recently, the incorporation of rotational mixing was thought to lead to
a much better agreement, the discovery that a significant fraction of
early B dwarfs are nitrogen-rich and intrinsically slowly rotating has cast
some doubts on the previous ideas. We argue that boron observations of early
type main sequence 
stars, as performed by Venn et al. (1996, 2002) and Morel et al. (2006, 2008),
have the potential to move towards a solution. 
Clearly, binary evolution needs to be  considered at the same time.
Finally, we may be facing the situation that still not all mixing processes
which can operate in massive main sequence stars have been described.
Amongst possible candidates is mixing due to magnetic processes, and
mixing induced by subsurface convection zones in hot massive stars.

\end{document}